\documentclass[12pt]{article}
\usepackage{latexsym}
\usepackage{amssymb}

\newcommand{\be}{\begin{equation}} \newcommand{\ee}{\end{equation}}
\newcommand{\bea}{\begin{eqnarray}} \newcommand{\eea}{\end{eqnarray}}

\def\am{Center For Theoretical Physics, Department of Physics\\ Texas A\&M University,
 College Station,TX 77843-4242,
USA}

\def\address#1{\begin{center}{ \it #1} \end{center}}
\def\author#1{\begin{center}{ \sc #1} \end{center}}
\def\title#1{\begin{center} {\Large #1 } \end{center}}
\def\Journal#1#2#3#4{{#1} {\bf #2}, #3 (#4)}

\def\EJC{{\em Eur.Phys.J} C}
\def\JHEP{\em JHEP}

\def\NPB{{\em Nucl. Phys.} B}
\def\PLB{{\em Phys. Lett.}  B}
\def\PRL{\em Phys. Rev. Lett.}
\def\PRD{{\em Phys. Rev.} D}

\def\TMP{\em Theor. Math. Phys}

\begin{document}

\begin{titlepage}

\title{Gravitational Forces in the Brane World}
\author{R. Arnowitt and J. Dent}
\address{\am}

\begin{abstract}
We consider the problem of gravitational forces between point particles on the branes in a Randall-Sundrum (R-S) two brane model with $S^1/Z_2$ symmetry.  Matter is assumed to produce a perturbation to the R-S vacuum metric and all the 5D Einstein equations are solved to linearized order (for arbitrary matter on both branes).  We show that while the gauge condition $h_{i5} = 0, i=0,1,2,3$ can always be achieved without brane bending, the condition $h_{55} = 0$ leads to large brane bending.  The static potential arising from the zero modes and the corrections due to the Kaluza-Klein (KK) modes are calculated.  Gravitational forces on the Planck ($y_1 = 0$) brane recover Newtonian physics with small KK corrections (in accord with other work).  However, forces on the TeV ($y_2$) brane due to particles on that brane are strongly distorted by large R-S exponentials.
\end{abstract}

\end{titlepage}

\section{Introduction}

Over the past two decades, the possible existence of extra dimensions has played an important role in much theoretical analysis.  Thus M-theory exists in 10 or 11 dimensions where six are compactified, characteristically on a Calabi-Yau (C-Y) manifold.  Examples of this are 10 dimensional (10D) intersecting D-brane models (see e.g.\cite{ibanez} and references therein), and 11 dimensional Horava-Witten (H-W) M-theory \cite{hw,hw2,witten,horava}.  In the latter case the eleventh dimension is compactified with $S^1/Z_2$ symmetry so that one can think of space as bounded by the two 10D orbifold planes at $y_1$ = 0 and $y_2 = \pi\rho$ with appropriate boundary conditions at $y_1$ and $y_2$ to enforce the $S^1/Z_2$ symmetry.  In H-W theory, only the gravity multiplet propagates in the bulk, while normal matter fields exist on the orbifold planes, the physical plane is conventionally chosen to be at $y_1 = 0$, while supersymmetry is broken on the distant $y_2$ plane \cite{horava}.  Since $\pi\rho$ is $\mathcal{O}$(10) times larger than the Calabi-Yau radius (which is $\approx 1/M_G$, where $M_G \backsimeq 3$x$10^{16}$GeV is the GUT mass) one can consider a 5D reduction of the theory with the effects of the C-Y space taken into account by moduli acting in the bulk, e.g. the C-Y volume moduli $\mathcal{V} = e^{\phi}$ being the simplest example.

The phenomenological 5D Randall-Sundrum (R-S) model \cite{rs,rs2} built on the same geometrical structure has received considerable analysis to see if such 5D models can reproduce standard 4D results.  The initial question examined was whether such a model could reproduce the 4D Friedmann-Robertson-Walker (FRW) cosmology\cite{binetruy,csaki,khoury,dewolfe,cline,cline2,enq,ellwanger,lukas,kobayashi,add}.  In order to do this, an additional scalar field in the bulk was added (analogous to the $\phi$ field in H-W) to stabilize the vacuum metric \cite{gw} with phenomenologically chosen potentials in the bulk and on the branes to fine tune the net cosmological constant on the brane to zero.  Matter on the branes is then treated as a perturbation (as appropriate for the Hubble expansion era) and it was seen that indeed the 4D FRW cosmology could be recovered.  A somewhat surprising result was that the same result could be obtained for H-W theory only in the radiation dominated era \cite{add}(at least for the tree level theory).  While here the potentials of H-W theory automatically guarantee the vanishing of the net cosmological constant without fine tuning, they are not free as in R-S models but are determined by the theory.  One finds that the phenomenologically chosen R-S potentials allow the brane separation to be determined by the non-relativistic matter density $\rho_{nr}$, something that does not occur in H-W theory\cite{add}.

A second question, which is the subject of this paper, is whether the 5D models give rise to the correct 4D Newtonian gravitational potential (in leading order).  To examine this, one considers point particles on the branes and calculates the gravitational forces between them.  There is a large literature on this subject as well and on calculations of corrections arising from the presence of an additional dimension\cite{lykken,tanaka,giddings,chung,dorca,deruelle,nojiri,callin}.  This question brings up an additional feature of the R-S model.  The vacuum metric of the R-S model has the form
\begin{equation}
ds^2 = e^{-2A(y)}\eta_{ij}dx^i dx^j + dy^2
\end{equation}
where A(y) is an increasing function of y.  (We use the notation i,j = 0,1,2,3; $\mu, \nu$ = 0,1,2,3,5 and $x^5 \equiv y$.)  Thus to account for the gauge hierarchy, the physical brane must be at $y_2$ (where $e^{-A(y_2)} \backsimeq 10^{-16}$) while in the H-W theory one can assume that the physical brane is at $y_1 = 0$.  Thus for R-S one needs to calculate the gravitational forces between the two particles at $y_2$ (though it is also interesting to see what forces the theory predicts between one particle at $y_1 = 0$ and one at $y_2 = \pi\rho$).  Previous analyses have only examined forces between particles on the $y_1 = 0$ brane\footnote{Ref.\cite{giddings} considers a single brane model with the brane displaced from the origin.  However this is different in that it does not have $S^1/Z_2$ boundary conditions imposed at $y=y_1$ and $y=y_2$.} and also neglect the effects of the Goldberger-Wise scalar stabilization field $\phi$.  For this case the function A(y) reduces to 
\begin{equation}
A(y) = \beta|y|\,\,\,;\,\,\, y_1 - \epsilon \leq y \leq y_2 - \epsilon\,\,\,;\,\,\,\epsilon > 0
\end{equation}
We examine here within this framework the general case of gravitational forces between particles on both branes as well as the size of the leading corrections due to the extra dimensions.  We find that the force on a particle on $y_1$ = 0 due to other particles on $y_1$ and $y_2$ has a leading Newtonian form (in accord with previous work) though the Newton constant $G_N$ is different for the two cases.  However, the Newtonian force on a particle at $y_2$ due to another particle at $y_2$ contains terms that grow exponentially with $y_2$ which leads to an unsatisfactory theory.

In carrying out these calculations, it is important to take careful account of ``brane bending'' effects.  Thus we assume that matter is added on the branes as a perturbation to the vacuum metric
\begin{equation}
ds^2 = e^{-2\beta y}(\eta_{ij}+ h_{ij})dx^i dx^j + h_{i5}dydx^i + (1 + h_{55})dy^2
\end{equation}
and then solve the Einstein equations to linear order in $h_{\mu\nu}$.  The diffeomorphisms of a 5D theory with $S^1/Z_2$ symmetry are those of $R^4$x$S^1$ which commute with $Z_2$.  This means that for the transformation
\begin{equation}
x^{\mu} \rightarrow x'^{\mu} + \xi^{\mu} \equiv x^{\mu}
\end{equation}
one has that $\xi^{5}$ vanishes at the orbifold points, $y_1$ and $y_2$:
\begin{equation}
\xi^5(x^i,y_1) = 0 = \xi^5(x^i,y_2)
\end{equation}
If one were to make a coordinate transformation with a non-vanishing $\xi^5$, then the branes become bent and this would create a complication when one imposes the $Z_2$ boundary constraint on the branes, leading to the so-called brane bending effects.  In previous analyses, the 5D Einstein equations were solved in Gaussian coordinates described by
\begin{equation}
h_{5\mu} = 0\,\,\,;\,\,\,\partial^{j}h_{ij} = 0 = \eta^{ij}h_{ij}
\end{equation}
In general, these cannot be achieved without brane bending occurring.  We give here an alternate analysis which avoids these complications by making only coordinate transformations that satisfy Eq.(5).

In Sec.2 we give the metric decomposition along with our gauge choices which will allow us to solve the Einstein equations in the presence of matter on the branes without introducing brane bending effects.  In Secs.3 and 4 we explicitly solve the bulk Einstein equations to first order in the metric perturbation in the static limit, and subject these solutions to the brane boundary conditions.  In Sec.4 we also find the poles of the transverse traceless piece of $h_{ij}$ and show how the Kaluza-Klein modes produce corrections to the leading static potential.  In Sec.5 we give the form of the Newtonian potential and show that Newton's constant differs depending on whether the gravitational force is due to particles on coincident or separate branes.  Conclusions are given in Sec.6 and the leading corrections to the Newtonian terms are discussed more fully in the Appendix.

\section{Coordinate Conditions}

Following 4D analyses, it is convenient to decompose the metric into its transverse and longitudinal parts according to the ADM prescription\footnote{This decomposition was first introduced in \cite{adm}. (Ref.\cite{adm2} is a more accessible recent reprint summarizing the ADM formalism.)  The generalization to 4-space is trivial except for the ambiguity in defining 1/$\Box^2$ in Minkowski space.  However, we will always be considering the static Newtonian limit here where $\Box^2$ $\rightarrow$ $\nabla^2$, though the size (and correct definition of) the higher order dynamical effect are also of interest.}:
\begin{equation}
h_{ij} = h_{ij}^{TT} + h_{ij}^{T} + h_{i,j} + h_{j,i}
\end{equation}
where $h_{ij}^{TT}$ is transverse and traceless\footnote{Four dimensional indices are raised and lowered with the Lorentz metric $\eta_{ij}$.  We use the notation $h_{i,j}$ $\equiv$ $\partial_{j}h_{i}$} ($\partial^{i}h_{ij}^{TT}$ $\equiv$ 0 $\equiv$ $h^{i}_{i}$) and $h_{ij}^{T}$ is transverse but (in general) possesses a trace:
\begin{equation}
\partial^{i}h_{ij}^{T} \equiv 0 \,\,\,;\,\,\, (h^{T})_{i}^{i} \equiv f^T \neq 0
\end{equation}
We also decompose $h_i$ into transverse and longitudinal parts
\begin{equation}
h_i = h_{i}^T + \frac{1}{2}h^{L}_{,i}\,\,\,;\,\,\, \partial^{i}h_{i}^{T} \equiv 0
\end{equation}
and can write
\begin{equation}
h_{ij}^{T} = \frac{1}{3}\pi_{ij}f^T \,\,\,;\,\,\, \pi_{ij} \equiv \eta_{ij} - O_{ij}
\end{equation}
where
\begin{equation}
O_{ij} \equiv \frac{\partial_i \partial_j}{\Box^2}
\end{equation}

One can express each of the subparts in Eq.(7) in terms of $h_{ij}$.  Thus taking the divergence of Eq.(7) gives
\begin{equation}
\partial^{i}h_{ij} = \partial_j \Box^2 h^L + \Box^2 h_{j}^{T}
\end{equation}
and
\begin{equation}
\partial^{i}\partial^{j}h_{ij} = (\Box^2)^2 h^L
\end{equation}
Thus
\begin{equation}
h^L_{,ij} = O_{ij}O_{kl}h^{kl}
\end{equation}
and
\begin{equation}
h^T_{i,j} = O^{k}_j h_{ik} - O_{ij}O_{kl}h^{kl}
\end{equation}
Taking the trace of Eq.(7) and using Eq.(14) determines $f^T$ to be
\begin{equation}
f^T = \pi^{ij}h_{ij}
\end{equation}
and since
\begin{equation}
h_{ij}^{TT} = h_{ij} - h^{T}_{ij} - h_{i,j} - h_{j,i}
\end{equation}
one has 
\begin{equation}
h_{ij}^{TT} = \pi_{ik}\pi_{jl}h^{kl} - \frac{1}{3}\pi_{ij}\pi_{kl}h^{kl}
\end{equation}

Our general metric with matter on the branes has the form
\begin{equation}
ds^2 = e^{-2A(y)}(\eta_{ij} + h_{ij})dx^i dx^j + h_{5i}dydx^i + (1 + h_{55})dy^2
\end{equation}
In order to discuss clearly the issues of brane bending, we assume here that there exists a frame with no brane bending (as e.g. is required in H-W theory) i.e. in this frame the 4D branes are orthogonal to the fifth dimension.  The vacuum metric of Eq.(1) is indeed a solution of the 5D Einstein equations obeying the $S^1/Z_2$ boundary conditions in this frame (as can be seen below Eqs.(37, 38)).  Since the perturbation due to matter, $h_{\mu\nu}$, in Eq.(19) is general, we can assume there is a choice of $h_{\mu\nu}$ that holds in a frame with no bending.  We now ask what coordinate conditions can be imposed to simplify the metric but remain in a frame with unbent branes.  We chose the coordinate condition
\begin{equation}
h_{5i} = 0\,\,\,;\,\,\,i = 0,1,2,3
\end{equation}
To verify that this can be achieved without any brane bending, we consider the infinitesimal gauge transformation
\begin{equation}
h'_{5i}(x) = h_{5i}(x) + \xi_{5,i} + \xi_{i,5} -2\Gamma_{5i}^{\alpha}\xi_{\alpha}
\end{equation}
where $\Gamma_{\mu\nu}^{\alpha}$ are the Christoffel symbols for the vacuum metric ($h_{\mu\nu}$ = 0).  Hence
\begin{equation}
h'_{5i}(x) = h_{5i}(x) + \xi_{5,i} + e^{-2A}(e^{2A}\xi_i)_{,5}
\end{equation}
Thus if in the initial gauge $h_{5i}$ were not zero, one can always set $h'_{5i}$ = 0 without brane bending e.g. by choosing $\xi_5$ = 0 and solving for $\xi_i$.  Eqs.(20) are not a complete set of coordinate conditions (being only four in number), and one may ask what is the remaining gauge freedom that maintains $\delta h_{5i}$ = 0.  From Eq.(22), this implies
\begin{equation}
0 =   \xi_{5,i} + e^{-2A}(e^{2A}\xi_i)_{,5}
\end{equation} 
Writing $\xi_i$ = $\xi_{i}^{T}$ + $\xi^{L}_{,i}$ where $\partial^{i}\xi_{i}^{T}$ $\equiv$ 0, one finds
\begin{eqnarray}
\omega_{i}^{T} = F_{i}^{T}(x^i)\\
\omega_5 = -(\omega^L)_{,5}
\end{eqnarray}
where $F_{i}^{T}$ is independent of y, and we have introduced the notation
\begin{equation}
\omega_{\mu} \equiv e^{2A(y)}\xi_{\mu}(x^i,y)
\end{equation}
It is understood in Eq.(25) that from Eq.(5) $\omega_5$($x^i$,$y_1$) = 0 = $\omega_5$($x^i$,$y_2$) to avoid brane bending, which consequently also constrains $(\omega^{L})_{,5}$.

The remaining gauge freedom, Eqs.(24, 25), allows a residual gauge freedom in $h_{ij}$ and $h_{55}$:
\begin{eqnarray}
\delta h_{ij} = \omega_{i,j} + \omega_{j,i} - 2A'e^{-2A}\omega_{5}\eta_{ij}\\
\delta h_{55} = 2(e^{-2A}\omega_5)_{,5}
\end{eqnarray}
where $A'(y)$ $\equiv$ d$A$/dy.  Decomposing $h_{ij}$ into its transverse and longitudinal parts one finds using Eqs.(24, 25) that
\begin{eqnarray}
\delta h_{ij}^{TT} = 0 \,\,\,;\,\,\, \delta h_{i}^{T} = \omega_{i}^{T}(x^i)\\
\delta f^T = 6A'e^{-2A}(\omega^{L}),_5 = -6A'e^{-2A}\omega_5\\
\delta(\Box^2 h^L) = 2\Box^2 \omega^L + 2A'e^{-2A}(\omega^{L})_{,5}
\end{eqnarray}
Since $\omega_5$($x^i$,$y_{\alpha}$) = 0 we see that $f^T$ is gauge invariant on the branes
\begin{equation}
\delta f^T(x^i,y_{\alpha}) = 0\,\,\,;\,\,\, \alpha = 1,2
\end{equation}
Further, since $\omega_5$($x^i$,y) is otherwise arbitrary, one may expand it around y = $y_{\alpha}$
\begin{equation}
\omega_5(x^i,y) = (y-y_{\alpha})\omega_{5}'(x^i,y_{\alpha}) + \frac{1}{2}(y-y_{\alpha})^{2}\omega_{5}''(x^i,y_{\alpha}) + ...
\end{equation}
so that
\begin{equation}
\delta(\partial_{5}f^T(x^i,y_{\alpha})) = -6A'e^{-2A}\omega'_5(x^i,y_{\alpha}) + ...
\end{equation}
Hence one may choose $\omega'_5(x^i,y_{\alpha})$ to set 
\begin{equation}
\partial_{5}f^T(x^i,y_{\alpha}) = 0
\end{equation}
which we will see below is a convenient further gauge choice.

\section{Einstein Equations}

The action for our system is
\begin{eqnarray}
S &=& \int d^5 x \sqrt{-^5 g}(-\frac{1}{2}M_{5}^{3}R + 6M_{5}^{3}\beta^2) \\\nonumber&+& \sum_{\alpha = 1,2}\int d^5 x \sqrt{-^4 g}(\mathcal{L}_{m_{\alpha}}+ (-1)^{\alpha}6M_{5}^{3}\beta)\delta(y-y_{\alpha})
\end{eqnarray}
where $M_5$ is the 5D Planck mass, $\mathcal{L}_{m_{\alpha}}$ are the Lagrangians for point particles on the branes $y_1$ = 0 and $y_2$ = $\pi\rho$, and we have fine tuned the bulk and brane cosmological constants so that the net cosmological constant is zero.  The vacuum equations of motion for the metric of Eq.(1) read
\begin{eqnarray}
\frac{1}{2}A'' &=& (-1)^{\alpha + 1}\beta \delta(y - y_{\alpha})\\
A'^2 &=& \beta^2
\end{eqnarray}
so that $A$ = $\beta y$ for $y_1 < y < y_2$ with $S^1$/$Z_2$ boundary conditions at the orbifold fixed points.  The linearized first order equations read
\begin{equation}
R_{\mu\nu}^{(1)} = -\frac{1}{M_{5}^3}\sum_{\alpha}(T_{\mu\nu}(y_{\alpha}) - \frac{1}{3}g_{\mu\nu}T(y_{\alpha}))\delta(y-y_{\alpha})
\end{equation}
where $T$ $\equiv$ $g^{\mu\nu}T_{\mu\nu}$ and $T_{\mu\nu}(y_{\alpha})$ are the 4D stress tensors for $\mathcal{L}_{m_{\alpha}}$.  Hence $T_{5i}$ = 0 = $T_{55}$.  We will consider here only the static gravitational forces and so only $T_{00}$ need to be taken non-zero\footnote{In obtaining Eq.(39), $R_{\mu\nu}^{(1)}$ is the first order part of $R_{\mu\nu}$ omitting terms proportional to the cosmological constant, since these terms are precisely canceled by the cosmological constant sources on the right hand side as a consequence of the zero'th order equations Eqs.(37, 38).}.

The 15 Einstein equations are then
\begin{eqnarray}
R_{5i}^{(1)} &=& 0\,\,\,;\,\,\, i = 0,1,2,3\\
R_{55}^{(1)} &=& -\frac{1}{3M_{5}^3}\sum_{\alpha}e^{2A}T_{00}(y_{\alpha})\delta(y-y_{\alpha})
\end{eqnarray}
and
\begin{equation}
R_{ij}^{(1)} = -\frac{1}{M_{5}^3}\sum_{\alpha}(T_{ij}(y_{\alpha}) -\frac{1}{3}\eta_{ij}e^{-2A}T)\delta(y-y_{\alpha})
\end{equation}
The 10 equations Eq.(42) can be decomposed as
\begin{eqnarray}
\eta^{ij}R_{ij}^{(1)} &=& -\frac{1}{3M_{5}^3}\sum_{\alpha}T_{00}(y_{\alpha})\delta(y-y_{\alpha})\\
\partial^{i}R_{ij}^{(1)} &=& -\frac{1}{M_{5}^3}\sum_{\alpha}(\partial^{i}T_{ij}(y_{\alpha}) + \frac{1}{3}\partial_{j}T_{00}(y_{\alpha}))\delta(y-y_{\alpha})
\end{eqnarray}
and
\begin{equation}
R_{ij}^{TT(1)} = -\frac{1}{M_{5}^3}\sum_{\alpha}T_{ij}^{TT}(y_{\alpha})\delta(y-y_{\alpha})
\end{equation}
Eqs.(43) and (44) together pick out the $''T''$ components and $''L''$ component of $R_{ij}$ while Eq.(45) picks out the $TT$ components.  In this section we solve Eqs.(40, 41, 43, 44).  Eq.(45) is discussed in Sec.4 below.

Eq.(40) reads
\begin{equation}
R_{5i}^{(1)} \equiv \frac{1}{2}\eta^{kl}\partial_{5}(\partial_{i}h_{kl} - \partial_{l}h_{ik}) + \frac{3}{2}A'\partial_{i}h_{55} = 0
\end{equation}
where $A' \equiv \partial_{y}A$.  In terms of the decomposition of Eqs.(7) and (9), Eq.(46) reduces to
\begin{equation}
-\partial_i\partial_5 f^T -3A'\partial_{i}h_{55} + \Box^{2}\partial_{5}h_{i}^{T} = 0
\end{equation}
where $\Box^2 \equiv \nabla^2-\partial_{0}^2$.  In the static approximation, $\Box^2 \rightarrow \nabla^2$, the $T$ part of Eq.(47) reads
\begin{equation}
\partial_5 h_{i}^T = 0
\end{equation}
which says that $h_{i}^T$ is independent of y.  We can thus use the gauge freedom of Eq.(29) to set $h_{i}^T$ to zero
\begin{equation}
h_{i}^T = 0
\end{equation}
The remaining longitudinal part of Eq.(47) yields
\begin{equation}
h_{55} = -\frac{1}{3A'}\partial_{5}f^T
\end{equation}
Note from Eqs.(28) and (30), the combination $3A'h_{55} + \partial_{5}f^T$ is gauge invariant.  However in the gauge of Eq.(35), one has that $h_{55}$ vanishes on the branes, i.e.
\begin{equation}
h_{55}(x^i,y_{\alpha}) = 0
\end{equation}
though in general it is non-zero off the branes.

Eqs.(41) and (43) allow us to determine $f^T$ and $h^L$.  One has for $R_{55}^{(1)}$
\begin{equation}
R_{55}^{(1)} = (\frac{1}{2}\partial_{5}^2 - A'\partial_5)\eta^{ij}h_{ij} + \frac{1}{2}e^{2A}\Box^2 h_{55} + 2A'\partial_{5}h_{55}
\end{equation}
and so Eq.(41) becomes, using Eq.(50)
\begin{equation}
(\frac{1}{2}\partial_{5}^2 - A'\partial_5)(\Box^{2}h^L - \frac{1}{3}f^T) - \frac{e^{2A}}{6A'}\Box^{2}\partial_{5}f^{T} = -\frac{1}{3M_{5}^3}\sum_{\alpha}e^{2A}T_{00}\delta(y-y_{\alpha})
\end{equation}
In arriving at Eq.(53) we have made use of the fact that \newline$A''\partial_{5}f^T$ $\sim$ $\delta(y-y_{\alpha})\partial_{5}f^T = 0$ in the gauge of Eq.(35).

The full $R_{ij}^{(1)}$ is
\begin{eqnarray}
e^{2A}R_{ij}^{(1)} &=& (\frac{1}{2}\partial_{5}^2 - 2A'\partial_5)h_{ij} - \frac{1}{2}A'\eta_{ij}\partial_{5}(\eta^{mk}h_{mk})\\\nonumber &+& \eta_{ij}h_{55}(A''-4A'^2) + \frac{1}{2}A'\eta_{ij}\partial_{5}h_{55}\\\nonumber &+& \frac{e^{2A}}{2}\partial_{i}\partial_{j}h_{55} + \frac{1}{2}e^{2A}\partial_{i}\partial_{j}(\eta^{mk}h_{mk})\\\nonumber &+& \frac{1}{2}e^{2A}\eta^{mk}(\partial_{k}\partial_{m}h_{ij} - \partial_{k}\partial_{i}h_{jm} - \partial_{j}\partial_{m}h_{ik})
\end{eqnarray}
Hence Eq.(43) becomes
\begin{eqnarray}\nonumber
(\frac{1}{2}\partial_{5}^2 - 4A'\partial_5)(\Box^{2}h^L - \frac{1}{3}f^T) - \frac{1}{6}\frac{e^{2A}}{A'}\Box^{2}\partial_{5}f^T + e^{2A}\Box^{2}f^T \\= -\frac{1}{3M_{5}^3}\sum_{\alpha}e^{2A}T_{00}(y_{\alpha})\delta(y-y_{\alpha})
\end{eqnarray}
Subtracting Eq.(55) from Eq.(53) determines $h^L$ in terms of $f^T$
\begin{equation}
\Box^{2}h^L = \frac{1}{3}f^T + \int_{0}^{y}dy'\frac{e^{2A}}{3A'}\Box^{2}f^T + \phi(x)
\end{equation}
where the function of integration $\phi(x)$ is independent of y.  However, from Eq.(31), one may set $\phi(x)$ to zero using a gauge transformation with $\omega^L(x^i)$.  Further, Eqs.(53) and (55) imply the same boundary conditions
\begin{equation}
\partial_{5}(\Box^{2}h^L - \frac{1}{3}f^T)\bigg|_{y=y_{\alpha}} = \frac{(-1)^{\alpha}}{3M_{5}^3}e^{2A(y_{\alpha})}T_{00}(y_{\alpha})
\end{equation}
Inserting Eq.(56) into Eq.(57) then gives
\begin{equation}
\Box^{2}f^{T}(x^i,y_{\alpha}) = \frac{(-1)^{\alpha}\beta}{M_{5}^3}T_{00}(y_{\alpha})
\end{equation}
where in our static approximation $\Box^2 \rightarrow \nabla^{2}$.  While $f^T$ is not gauge invariant in the bulk, we saw that it was gauge invariant on the branes, which is why its value on each brane is determined by the physical quantities $T_{00}(y_{\alpha})$.

If one now inserts Eq.(56) back into Eqs.(53) and (55), one sees that these equations are identically satisfied and so Eqs.(53) and (56) have no further content.  Thus rather than determining $h^L$ and $f^T$ separately, Eqs.(53) and (55) determine only the gauge invariant combination Eq.(56).

To check the solution of Eq.(44), we first note that $\partial^{i}T_{ij}$ is second order and may be neglected.  Using Eq.(54) and the coordinate conditions Eqs.(49) and (51), and the fact that
\begin{equation}
A''h_{55}  \sim \delta(y-y_{\alpha})h_{55} = 0
\end{equation}
one sees that Eq.(44) reduces to
\begin{eqnarray}\nonumber
(\frac{1}{2}\partial_{5}^2 - \frac{5A'}{2}\partial_5)(\Box^{2}h^L - \frac{1}{3}f^T) &-& \frac{1}{6}\frac{e^{2A}}{A'}\Box^{2}\partial_{5}f^T + \frac{1}{2}e^{2A}\Box^{2}f^T = \\&-&\frac{1}{3M_{5}^3}T_{00}(y_{\alpha})\delta(y-y_{\alpha})
\end{eqnarray}
The boundary conditions implied by the right hand side of Eq.(60) are thus identical to Eq.(57), and inserting Eq.(56) one sees that Eq.(60) is identically satisfied. 

We will see in the following section that the remaining Einstein equations, Eq.(45) uniquely determines $h_{ij}^{TT}$ so that we have found solutions to all the Einstein equations.  The undetermined function, $f^{T}(x^i,y)$ off the branes, is the remaining gauge freedom.  However, note one cannot set $h_{55}$ = 0 everywhere (as is conventionally done in other analyses) as Eq.(50) would then imply $f^T$ is constant in y, which would be inconsistent with the boundary conditions Eq.(58) (which are in fact gauge invariant).  We will see that Eq.(58) contributes a significant term to the static gravitational potential.

\section{Solution For $h_{ij}^{TT}$}

The remaining Einstein field equations, Eq.(45), can be obtained by taking the $TT$ part of Eq.(54).  We find
\begin{equation}
(\frac{1}{2}\partial_{5}^{2} -2A'\partial_{5} + \frac{1}{2}e^{2A}\Box^2)h_{ij}^{TT} = -\frac{e^{2A}}{M_{5}^3}\sum_{\alpha}T_{ij}^{TT}(y_{\alpha})\delta(y-y_{\alpha})
\end{equation}
and hence $h_{ij}^{TT}$ obeys the boundary conditions
\begin{equation}
\partial_5 h_{ij}^{TT}\bigg|_{y=y_{\alpha}} = (-1)^{\alpha}\frac{e^{2A}}{M_{5}^3}T_{ij}^{TT}(y_{\alpha})\,\,\,;\,\,\, \alpha = 1,2
\end{equation}
The static potential is obtained from $h_{00}(x,y_{\alpha})$ where
\begin{equation}
h_{00}(x^i,y_{\alpha}) = h_{00}^{TT}(x^i,y_{\alpha}) - \frac{1}{3}f^T(x^i,y_{\alpha})
\end{equation}
The corresponding source is then
\begin{equation}
T_{00}^{TT} = \pi_{0}^{\rho}\pi_{0}^{\sigma}T_{\rho\sigma} - \frac{1}{3}\pi_{00}\pi_{\rho}^{\sigma}T_{\rho}^{\sigma}
\end{equation}
which in the static limit reduces to
\begin{equation}
T_{00}^{TT} = \frac{2}{3}T_{00}
\end{equation}

To solve Eq.(61) we Fourier analyse $h_{ij}^{TT}$
\begin{equation}
h_{ij}^{TT}(x^i,y_{\alpha}) = \int d^{4}p e^{ipx}h_{ij}^{TT}(p^i,y)
\end{equation}
In the bulk then $h_{ij}^{TT}(p,y)$ obeys
\begin{equation}
(\frac{1}{2}\partial_{5}^2 - 2A'\partial_5 + \frac{1}{2}e^{2A}m^2)h_{ij}^{TT}(p,y) = 0
\end{equation}
where $m^2 \equiv -p^2 = p_{o}^2 - \vec{p}^2$.  The solutions of Eq.(67) are Bessel and Neumann functions
\begin{equation}
h_{ij}^{TT}(p^i,y) = e^{2\beta y}[A_{ij}(p)J_{2}(\xi) + B_{ij}(p)N_{2}(\xi)]
\end{equation}
where
\begin{equation}
\xi(y) = \frac{m}{\beta}e^{\beta y}
\end{equation}
and $m/\beta$ is short hand for $(m^2/\beta^{2})^{1/2}$.  The boundary conditions Eq.(62) determine $A_{ij}$ and $B_{ij}$.  One finds on the branes
\begin{equation}
h_{00}^{TT}(p;y_1) = -\frac{2}{3\beta M_{5}^3}[\frac{N_{11}(\xi_1,\xi_2)}{D}T_{00}(y_1) + \frac{N_{12}(\xi_1,\xi_2)}{D}T_{00}(y_2)]
\end{equation}
where 
\begin{eqnarray}
D \equiv \frac{N_1(\xi_1)}{J_1(\xi_1)} - \frac{N_1(\xi_2)}{J_1(\xi_2)} \\
\xi_1 = \frac{m}{\beta}\,\,\,; \,\,\,\xi_2 = \frac{m}{\beta}e^{\beta y_2}\\
m^2 \equiv -p^2 = (p^{0})^2 - \vec{p}^2\\
N_{11} \equiv \frac{J_2(\xi_1)}{\xi_1 J_1(\xi_1)}\left[\frac{N_2(\xi_1)}{J_2(\xi_1)} - \frac{N_1(\xi_2)}{J_1(\xi_2)}\right]\\
N_{12} \equiv \frac{J_2(\xi_1)}{\xi_2 J_1(\xi_2)}\left[\frac{N_2(\xi_1)}{J_2(\xi_1)} - \frac{N_1(\xi_1)}{J_1(\xi_1)}\right]
\end{eqnarray}
and
\begin{equation}
h_{00}^{TT}(p;y_2) = -\frac{2e^{2\beta y_2}}{3\beta M_{5}^3}[\frac{N_{21}(\xi_1,\xi_2)}{D}T_{00}(y_1) + \frac{N_{22}(\xi_1,\xi_2)}{D}T_{00}(y_2)]
\end{equation}
where 
\begin{equation}
N_{21}(\xi_1,\xi_2) = N_{12}(\xi_2,\xi_1)\,\,\,;\,\,\, N_{22}(\xi_1,\xi_2) = N_{11}(\xi_2,\xi_1)
\end{equation}
It is useful to consider $h_{00}^{TT}(p,y_{\alpha})$ as a function of a complex variable $z = m^2$.  In looking at the analytic behavior in z, the logarithmic branch cuts in the Neumann functions cancel in the differences such as $N_1(\xi_1)/J_1(\xi_1) - N_1(\xi_2)/J_1(\xi_2)$.  Poles can arise from a number of sources.  Thus in $N_{11}$, a pole might occur at the zeros of $J_1(\xi_1)$, but this is actually canceled by the zeros in $J_1(\xi_1)$ appearing in $D$.  Similarly the zeros of $J_1(\xi_2)$ in $N_{11}$ are canceled.  Thus the only poles that occur are the pole at $m^2 = 0$ (arising e.g. from $N_2/J_2 \sim 1/m^4$ in $N_{11}$) and when $D$ vanishes, i.e., at
\begin{equation}
D(m_{n}^2) = 0 = \frac{N_1(\xi_n)}{J_1(\xi_n)} - \frac{N_1(\xi_n e^{\beta y_2})}{J_1(\xi_n e^{\beta y_2})}
\end{equation}
where $\xi_n \equiv (m_{n}^2/\beta^2)^{1/2}$.  To find the residue at $m_{n}^2$, we expand $D(m_{n}^2)$ around the pole position
\begin{equation}
D(m^2) = \frac{\partial D(m^2)}{\partial m^2}\bigg|_{m_{n}^2}(m^2 - m_{n}^2) + \ldots
\end{equation}
and differentiating the Bessel and Neumann functions in D one has
\begin{eqnarray}
\frac{\partial D(m^2)}{\partial m^2} &=& \frac{1}{2\beta m}\frac{1}{J_{1}^2(\xi_1)}[J_1(\xi_1)N_{1}'(\xi_1) - N_1(\xi_1)J_{1}'(\xi_1)]\\\nonumber &-& \frac{e^{\beta y_2}}{2\beta m}\frac{1}{J_{1}^2(\xi_2)}[J_1(\xi_2)N_{1}'(\xi_2) - N_1(\xi_2)J_{1}'(\xi_2)]
\end{eqnarray}
Using the two Bessel function Wronskian identities
\begin{equation}
J_1 N_{1}' - N_{1}J_{1}' = \frac{2}{\pi\xi} = N_1 J_2 - J_1 N_2
\end{equation}
Eq.(79) reduces to
\begin{equation}
D(m^2) = (\frac{1}{\pi m^2}[\frac{1}{J_{1}^2(\xi_1)} - \frac{1}{J_{1}^2(\xi_2)}])\bigg|_{m_n}(m^2 - m_{n}^2) + \ldots
\end{equation}
To obtain the residue at the pole, we need also the numerator $N_{11}$ evaluated at $m^2 = m_{n}^2$
\begin{equation}
N_{11} = \frac{1}{\xi_1}\frac{1}{J_{1}^{2}(\xi_1)}[N_{2}(\xi_1)J_1(\xi_1) - N_{1}(\xi_1)J_2(\xi_1)]\bigg|_{m_n}
\end{equation}
and using Eq.(81) this reduces to
\begin{equation}
N_{11}(m_{n}^2) = -\frac{2\beta^2}{\pi m_{n}^2}\frac{1}{J_{1}^2(m_{n}^2/\beta^2)}
\end{equation}
Hence the residue at $m_{n}^2$ is simply
\begin{equation}
R_n(m_{n}^2) = -\frac{2}{3\beta M_{5}^3}\frac{N_{11}}{D}\bigg|_{m_n} = \frac{4\beta}{3M_{5}^3}\frac{1}{1-\frac{J_{1}^{2}(m_n/\beta)}{J_{1}^{2}((m_n/\beta)e^{\beta y_2})}}
\end{equation}
The residue at $m^2 = 0$ can be obtained by examining the limit when both $\xi_1$ and $\xi_2$ approach zero in $N_{11}(\xi_1,\xi_2)/D$.  One finds
\begin{equation}
R_0 = -\frac{4\beta}{3M_{5}^3}\frac{1}{1-e^{-2\beta y_2}}
\end{equation}
where for the Randall-Sundrum model one may neglect the $e^{-2\beta y_2} \approx 10^{-32}$ in the denominator.  A similar analysis to the above holds for the other three terms in Eqs.(70) and (76).

One may now cast the results for $h_{00}^{TT}(z,y_{\alpha})$ in a more convenient form.  Using the asymptotic forms of Bessel and Neumann functions, one can see that $h_{00}^{TT}(z,y_{\alpha})$ falls like 1/z for $|z|$ on a large circle.  Hence integrating
\begin{equation}
g(z) \equiv \frac{h_{00}^{TT}(z,y_{\alpha})}{z-m^2}
\end{equation}
over a large circle in the complex plane we can express $h_{00}^{TT}(m^2,y_{\alpha})$ as a sum of poles with the residues calculated above.  One has then
\begin{eqnarray}
&&h_{00}^{TT}(y_1) = -\frac{4\beta}{3m^{2}M_{5}^3}[T_{00}(y_1) + e^{-2\beta y_2}T_{00}(y_2)]\\\nonumber
&&+ \frac{4\beta}{M_{5}^3}\sum_{m_n}\frac{1}{m^2 - m_{n}^2}\left(\frac{J_{2}^{2}(\xi_2)}{J_{2}^{2}(\xi_2) - J_{1}^{2}(\xi_1)}\right)[T_{00}(y_1) + \frac{e^{-\beta y_2}J_{1}(\xi_1)}{J_{1}(\xi_2)}T_{00}(y_2)]\bigg|_{m=m_n}
\end{eqnarray}
and
\begin{eqnarray}
&&h_{00}^{TT}(y_2) = -\frac{4\beta}{3m^{2}M_{5}^3}[T_{00}(y_1) + e^{-2\beta y_2}T_{00}(y_2)]\\\nonumber
&&+ \frac{4\beta}{M_{5}^3}\sum_{m_n}\frac{1}{m^2 - m_{n}^2}\left(\frac{J_{1}^{2}(\xi_1)}{J_{1}^{2}(\xi_2) - J_{1}^{2}(\xi_1)}\right)[T_{00}(y_2) + \frac{e^{\beta y_2}J_{1}(\xi_2)}{J_{1}(\xi_1)}T_{00}(y_1)]\bigg|_{m=m_n}
\end{eqnarray}
The first term of Eqs.(88) and (89) contributes to the Newtonian potential (since $m^2$ = $-p^{\,2} \rightarrow \vec{p}^2$ in the static limit) while the other term gives 5D corrections to the Newtonian theory.

While Eq.(78) is a transcendental equation, one can obtain the positions of the poles analytically in certain limits.  Thus if $\xi_n$ = $m_n/\beta \ll 1$ but $\xi_n e^{\beta y_2} \gg 1$ (i.e. $\xi_n \gg 10^{-16}$) then inserting in the Bessel function asymptotic forms in the ratios of Eq.(78) gives
\begin{equation}
tan(\xi_n e^{\beta y_2} - \frac{3\pi}{4}) \cong -\frac{4}{\pi}(\frac{1}{\xi_n})^2
\end{equation}
which can be solved by iteration to give
\begin{equation}
\frac{m_{n}^2}{\beta^2} \cong [(n+\frac{5}{4})\pi + \epsilon_n]^{2}e^{-2\beta y_2} \,\,\,;\,\,\, \xi_n \ll 1\,\,,\,\, \xi_n e^{\beta y_2} \gg 1
\end{equation}
where
\begin{equation}
\epsilon_n \cong \frac{\pi}{4}[(n+\frac{5}{4})\pi e^{-\beta y_2}]^2
\end{equation}
(We have included the first order correction $\epsilon_n$ as in some expressions the leading term can cancel out).  The residues at the poles can then be calculated in this limit.  Thus for the $T_{00}(y_1)$ term one finds for
\begin{equation}
R_n = -\frac{2}{3\beta M_{5}^3}\frac{N_{11}(\xi_1,\xi_2)}{D}\bigg|_{m=m_n}
\end{equation}
the result
\begin{equation}
R_n = -\frac{2\pi}{3 M_{5}^3}m_n e^{-\beta y_2}
\end{equation}
and the contribution to the scalar potential is
\begin{equation}
h_{00}^{TT} = -\frac{2\pi}{3 M_{5}^3}\sum_{n=1}\frac{m_n e^{-\beta y_2}}{m^2 - m_{n}^2}T_{00}(y_1)
\end{equation}
Since the poles are very dense
\begin{equation}
\Delta m_n \equiv m_{n+1} - m_n = \beta\pi e^{-\beta y_2}
\end{equation}
one can approximate Eq.(95) by converting the sum to an integral ($m^2 = -p^2 = -\vec{p}^{\,2}$ in the static limit)
\begin{equation}
h_{00}^{TT} \cong \frac{2\pi}{3 M_{5}^3}\int_{0}^{\beta}dm_{n}\frac{m_n}{\vec{p}^2 + m_{n}^2}T_{00}(y_1)
\end{equation}
where we have cut off the integral at $\beta$ since $\xi_1 = m_n/\beta \lesssim 1$.  Returning to coordinate space this yields
\begin{equation}
h_{00}^{TT}(\vec{r}) = \frac{2}{3\beta M_{5}^3}\frac{m_0}{4\pi r^3}\int_{0}^{\beta r}d\alpha e^{-\alpha}
\end{equation}
where $m_0$ is the mass of $T_{00}(y_1)$.  For $r \gg 1/\beta$, i.e. for distances large compared to the warping parameter $1/\beta$, this is a $1/r^3$ correction to the leading Newtonian potential.

Eventually, for sufficiently large n, $\xi_n$ becomes large (i.e. $n \gtrsim 10^{16}$), and the poles from Eq.(78) occur at
\begin{equation}
m_n = \frac{n\pi\beta}{e^{\beta y_2} - 1} \cong n\pi\beta e^{-\beta y_2} \,\,\,;\,\,\, \xi_n \gg 1
\end{equation}
Then one finds for this contribution
\begin{equation}
h_{00}^{TT} = -\frac{4\beta}{3 M_{5}^3}\sum_{n}\frac{e^{-\beta y_2}}{m^2 - m_{n}^2}T_{00}(y_1)
\end{equation}
or in the continuum approximation
\begin{equation}
h_{00}^{TT} \cong \frac{4\beta}{3 M_{5}^3}\int_{\beta}^{\infty}\frac{dm_n}{\vec{p}^2 + m_{n}^2}T_{00}(y_1)
\end{equation}
In coordinate space one finds
\begin{equation}
h_{00}^{TT}(r) = \frac{m_0}{3 M_{5}^3}\int_{\beta}^{\infty}\frac{dm_n e^{-m_{n}r}}{r}T_{00}(y_1)
\end{equation}
and in the limit $r \ll 1/\beta$ one finds a $1/r^2$ correction to the Newtonian potential
\begin{equation}
h_{00}^{TT}(r) = \frac{m_0}{3M_{5}^3}\frac{1}{r^2}
\end{equation}
Eqs.(98) and (103) agree with results obtained in \cite{callin}(although there, a fine tuning of matter is needed on the second brane in order to get a consisitent solution of the Einstein equations).

One may carry out a similar analysis of the other three terms in Eqs.(70) and (76) ($N_{12}/D$, $N_{21}/D$, and $N_{22}/D$) and these results will be discussed further in the Appendix.  In Eq.(97) it is conventional to extend the integral down to $m_n$ = 0, and think of the continuum of poles as reaching down to $m^2 = 0$ without a gap.  Actually, as can be seen from Eq.(91), the first discrete pole occurs at
\begin{equation}
m_n \cong (\frac{9}{4}\pi\beta)e^{-\beta y_2}
\end{equation} 
The size of the gap depends on the model.  Thus for Randall-Sundrum one has
\begin{equation}
m_1 \approx (\frac{9}{4}\pi\beta)e^{-\beta y_2} \approx (10^{19}GeV)(10^{-16}) = 1TeV
\end{equation}
since
\begin{equation}
\beta \approx M_{Pl}
\end{equation}
On the Planck brane $y_1 = 0$, a TeV of energy is negligible (since masses are of order $M_{Pl}$).  On the TeV brane $y_2$ however, it is sometimes argued that one should not consider phenomena $\gtrsim 1$TeV.  In this case one would neglect the Kaluza-Klein modes.

\section{Newtonian Potential}
The static Newtonian potential is the 1/r terms of $h_{00}(x^i,y_{\alpha})$ of Eq.(63).  These arise from the poles in momentum space at $m^2 = 0$.  As discussed in Sec.4, these poles occur in $h_{00}^{TT}$ from the fact that the numerator functions $N_{ij}$ go as $N_{ij} \sim 1/m^4$ as $m^2 \rightarrow 0$ due to the $N_{2}(\xi_{\alpha}), \alpha = 1,2$ terms, while the denominator function goes as $D \sim 1/m^2$ leading to a net 1/$m^2$ term for small $m^2$.  As seen from Eq.(58), $f^T(x^i,y_{\alpha})$ is totally a $1/m^2$ term in momentum space.  One can thus pick out the $m^2 = 0$ pole contributions on the two branes
\begin{equation}
h_{00}^{N}(y_1) = -\frac{4\beta}{3M_{5}^3}\frac{1}{m^2}[T_{00}(y_1) + e^{-2\beta y_2}T_{00}(y_2)] + \frac{\beta}{3M_{5}^3}\frac{1}{m^2}T_{00}(y_1)
\end{equation}
and
\begin{equation}
h_{00}^{N}(y_2) =  -\frac{4\beta}{3M_{5}^3}\frac{1}{m^2}[T_{00}(y_1) + e^{-2\beta y_2}T_{00}(y_2)] - \frac{\beta}{3M_{5}^3}\frac{1}{m^2}T_{00}(y_2)
\end{equation}
In Eqs.(107) and (108) the first bracket is from $h_{00}^{TT}$ and the second is from $f^T$.  The stress tensor $T_{ij}$ arising from the matter Lagrangian $\mathcal{L}_m$ is
\begin{equation}
T^{ij} = \frac{1}{\sqrt{-g}}\frac{\delta\mathcal{L}_m}{\delta g_{ij}}
\end{equation}
where for a point particle on brane $y_{\alpha}$
\begin{equation}
\mathcal{L}_{m_{\alpha}} = m_{0}\int d\tau u^i u^j g_{ij}(x^i,y_{\alpha})\delta^4(x^i - x^i(\tau))
\end{equation}
and $u^i = dx^i/d\tau$ with $d\tau^2 = -g_{ij}dx^{i}dx^{j}$.  For our metric, $g_{ij} = e^{-2A}\hat{g_{ij}}$ where in the linearized approximation $\hat{g_{ij}} = \eta_{ij} + h_{ij}$.  Thus defining
\begin{equation}
d\hat{\tau}^2 = \hat{g_{ij}}dx^i dx^j \,\,\,;\,\,\, \hat{u}^i = \frac{dx^i}{d\hat{\tau}}
\end{equation}
$\mathcal{L}_m$ reduces to 
\begin{equation}
\mathcal{L}_{m_{\alpha}} = \bar{m}_{\alpha}(y_{\alpha})\int d\hat{\tau}\hat{u}^i\hat{u}^j \hat{g_{ij}}\delta^{4}(x^i - x^i(\hat{\tau}))
\end{equation}
where
\begin{equation}
\bar{m}(y) = e^{-A(y)}m_0
\end{equation}
showing the usual result that if $m_0$ is of Planck size, the effective mass seen on the TeV brane $y_2$ will be of TeV size.  The Lagrangian of Eq.(112) will then correctly give rise to the (linearized) geodesic equation governed by $\hat{g_{ij}} = \eta_{ij} + h_{ij}$.

Returning to Eq.(109), the stress tensor is
\begin{equation}
T^{ij} = \frac{1}{\sqrt{-g}}m_0\int d\tau u^i u^j \delta^4(x^i - x^i(\tau))
\end{equation}
and in the static approximation,
\begin{equation}
u^0 \cong e^A \,\,\,;\,\,\, u^i \cong 0
\end{equation}
one has
\begin{equation}
T^{00} = e^{5A}m_0 \delta^{3}(r-r(t))
\end{equation}
so that
\begin{equation}
T_{00}(y_{\alpha}) = e^{2A(y_{\alpha})}\bar{m}(y_{\alpha}) \delta^{3}(r-r(t))
\end{equation}

The interaction potential between the two particles may be defined by
\begin{equation}
V = -\int d^{3}r\mathcal{L}_{m\,\, int}
\end{equation}
where the total $\mathcal{L}_m$ is
\begin{equation}
\mathcal{L}_m = \sum_{\alpha}m_{0_{\alpha}}\int d\tau u^i u^j e^{-2A(y_{\alpha})}(\eta_{ij} + h_{ij}(y_{\alpha}))\delta^{4}(x-x_{\alpha}(\tau))
\end{equation}
Hence in the static limit
\begin{equation}
-\int d^{3}r\mathcal{L}_m = -\sum_{\alpha}m_{0_{\alpha}}\int d\tau (\frac{dx^{0}_{\alpha}}{d\tau})^2 e^{-2A(y_{\alpha})}(\eta_{00} + h_{00}(y_{\alpha}))\delta^{4}(x-x_{\alpha}(\tau))
\end{equation}
Since $u^{0}_{\alpha}d\tau  = dx^{0}_{\alpha}$ and
\begin{equation}
\frac{dx^{0}_{\alpha}}{d\tau} \cong \frac{1}{\sqrt{-g_{00}}} = \frac{e^A}{(-\eta_{00} - h_{00})^{1/2}}
\end{equation}
one has
\begin{equation}
-\int d^{3}r\mathcal{L}_m = \sum_{\alpha}\bar{m}_{\alpha}(-\eta_{00} - h_{00})^{1/2}
\end{equation}
Expanding to first order gives for the interaction potential energy
\begin{equation}
V = -\frac{1}{2}\sum_{\alpha}\bar{m}_{\alpha}h_{00}(x^{0}_{\alpha},y_{\alpha})
\end{equation}
Inserting Eq.(107) and (108) and returning to coordinate space ($m^2 = -\vec{p}^2$) one gets for the Planck brane the contribution
\begin{equation}
V(y_1) = -\frac{\beta}{8\pi M_{5}^3}\frac{1}{r}[\bar{m}_1\bar{m}_1' + \frac{4}{3}\bar{m}_1\bar{m}_2]
\end{equation}
where $\bar{m}_1$ is the mass of a second particle on the Planck brane, $\bar{m}_2$ a mass of a particle on the TeV brane ($\bar{m}_2 = e^{-\beta y_2}m_{20}$), and $r$ is the 3D distance between the particles.  Note that the fact that $\bar{m}_2$ is separated by additional distance in the fifth dimension ($y_2 - y_1 = \pi\rho$) does not enter in $r$.

We see from Eq.(124) that if the two particles are on the Planck brane, Eq.(126) correctly reproduces the Newtonian force law with
\begin{equation}
G_N \equiv \frac{\beta}{8\pi M_{5}^3}
\end{equation}
(the conventional value for the Newton constant in Randall-Sundrum theory).  The $f^T$ contribution correctly changes the 4/3 factor in the first term of Eq.(107) to 1.  However, if one particle is on the TeV brane, the Newton constant is modified by an extra factor of 4/3, since the $f^T$ factor does not contribute. (The fact that matter on the TeV brane changes gravitational effects seen on the Planck brane has previously been noted in \cite{tanaka} in a different connection.)

For the potential energy seen on the TeV brane we use Eq.(108) in Eq.(123).  One finds now from Eq.(117) that
\begin{eqnarray}\nonumber
V(y_2) = -\frac{4\beta}{3 M_{5}^3}\frac{1}{8\pi r}[\bar{m}_2\bar{m}_1 + \bar{m}_2\bar{m}_2'] \\- \frac{\beta}{3 M_{5}^3}\frac{1}{8\pi r}\bar{m}_2\bar{m}_2'e^{2\beta y_2}
\end{eqnarray}
where $\bar{m}_2'$ is a second particle on the $y_2$ brane.  The interaction energy between $\bar{m}_2$ and $\bar{m}_1$ particles is as before as is the Newtonian potential between two particles on the TeV brane, $m_2$ and $m_2'$, arising from $h_{00}^{TT}$ (aside from the peculiar 4/3 factor).  However, the $f^T$ term gives an additional contribution to $V(y_2)$ scaled by $e^{2A(y_2)}$ (the factor from Eq.(117)) which would produce an anomolously large additional contribution. (Recall $e^{\beta y_2} \approx 10^{16}$ in the Randall-Sundrum model to account for the gauge hierarchy problem!)  Thus the theory does not appear to give sensible results on the TeV brane.

\section{Conclusions}

In this paper we have examined the gravitational forces between point particles in the static limit in the two brane Randall-Sundrum model.  In contrast to previous analyses, we have chosen gauge conditions (coordinate frames) to solve the field equations that maintain the $S^1/Z_2$ boundary conditions, and hence produce no brane bending effects, and we also examine forces between particles on both branes, not just the $y_1 = 0$ brane.  A convenient technique for solving the field equations is to introduce for the 4D generalization of the ADM decomposition \cite{adm,adm2} for the metric perturbation
\begin{equation}
h_{ij} = h_{ij}^{TT} + h_{ij}^T + h_{i,j} + h_{j,i}
\end{equation}
where $\partial^{i}h_{ij}^{TT} = 0 = \partial^{i}h_{ij}^T$, $\eta^{ij}h_{ij}^{TT} = 0$, and $\eta^{ij}h_{ij}^{T} \equiv f^T \neq 0$.  The $h_{ij}^{TT}$ contain the Kaluza-Klein modes while both the $h_{ij}^{TT}$ and $h^{T}_{ij}$ contribute to the static Newtonian potential (with pole at $\vec{p}^{\,2}$ = 0 in momentum space).  One finds that a particle on the $y_1$ = 0 brane sees a Newtonian force from another particle on either the $y_1$ brane or the $y_2 = \pi\rho$ brane but with different Newtonian constants: $G_N = \beta/8\pi M_{5}^3$ and $G_N = \beta/6\pi M_{5}^3$ respectively.  The difference arises from the fact that the $f^T$ component of $h_{00}$ enters with opposite sign for $y_1$ and $y_2$ particles, as seen in Eqs.(58) and (63).  (The fact that matter at $y_2$ effects matter at $y_1$ differently from other matter at $y_1$ was also noted in \cite{tanaka} in another connection.)  Note that the $f^T$ contribution is precisely what is needed to give the conventional value $G_N = \beta/8\pi M_{5}^3$ on the $y_1$ brane.

A curious feature of the potential Eq.(124) is that the force depends only on the 3D distance, and is independent of any $y$ separation.  It would be interesting to see if this produces any causal questions i.e. if one jiggled the mass on $y_2$, how long does it take for the effect to become noticeable at the $y_1$ particle, a question involving dynamical rather than static solutions.

A more serious problem is the force seen by two particles on the TeV brane $y_2$.  One sees from Eq.(126) that there is an attractive term arising from the $f^T$ contribution which is $\mathcal{O}(e^{2\beta y_2}) \approx (10^{16})^2$ larger than the normal gravity and this occurs \emph{after} one has correctly rescaled the $y_2$ masses to TeV size (as one normally does in the RS model).  Thus one does not recover normal Newtonian gravitation in the static limit on the TeV brane.  All analyses up to now have neglected the Goldberger-Wise scalar field.  Including it in might in some way cancel out the anomolous $e^{2\beta y_2}$ factor in Eq.(126).  The analysis including the scalar field is much more complicated than the calculation given here.  Preliminarily we find equations for $f^T$ at the brane positions $y_{\alpha}$ to be given in terms of the scalar field, $\phi$, by
\begin{equation}
\Box^2 f^T(y_{\alpha}) = \frac{(-1)^{\alpha}}{M_{5}^3}A'T_{00} - \frac{2e^{-2A}}{M_{5}^3}[\phi_{o}''\delta\phi - \phi_{o}'\delta\phi']
\end{equation}
Where $\delta\phi$ is the pertubation of the scalar field about the vacuum value $\phi_o$.  The perturbation $\delta\phi$ satisfies a complicated equation of motion given in terms of the bulk potential $V(\phi_{o})$ and its derivatives with respect to $\phi$ as well as the brane potential $V_{\alpha}(\phi_o)$:
\begin{eqnarray}\nonumber
&&e^{2A}\Box^2 \delta\phi - 4A'\delta\phi' + \delta\phi'' - \delta\phi V''(\phi_o) - V'(\phi_o)h_{55} \\\nonumber && -\frac{1}{2}\phi_{o}'h_{55}' + \frac{1}{2}\phi_{o}'\partial_y(\Box^2 h^L + f^T) \\ &&= \sum_{\alpha}\delta(y-y_{\alpha})(\frac{1}{2}V_{\alpha}'(\phi_o)h_{55} + V_{\alpha}''(\phi_o)\delta\phi)
\end{eqnarray}
which is not decoupled from $f^T$.  The remaining equations for $h^L$ and $h_{55}$ are 
\begin{eqnarray}
h_{55} &=& -\frac{1}{3A'}\partial_y f^T - \frac{2\phi_{o}'}{3A'M_{5}^3}\delta\phi\\\nonumber
3A'\partial_y(\Box^2 h^L +f^T)&=& \frac{2}{M_{5}^3}V(\phi_o)h_{55} + \frac{2}{M_{5}^3}V'(\phi_o)\delta\phi\\&& -\frac{2\phi_{o}'}{M_{5}^3}\delta\phi' - e^{2A}\Box^2f^T
\end{eqnarray}
At present it is not obvious that once the solution of this system is found (and then substituted into Eq.(128)) that this will rid the system of the exponentially large Newtonian force felt on the brane at $y_2$.  

We briefly compare our analysis with some of the previous calculations for the static gravitational potential.  In Ref.\cite{tanaka} it is assumed that in Gaussian coordiinates
\begin{equation}
h_{\mu 5} = 0 \,\,\,;\,\,\, \mu = 0,1,2,3,5
\end{equation}
there is no brane bending, and brane bending occurs only when one adds the coordinate conditions
\begin{equation}
\partial^{i}h_{ij} = 0 = h^{i}_{i} \,\,\,;\,\,\, i,j = 0,1,2,3
\end{equation}
One can easily check, however, that the extra condition $h_{55} = 0$ of Eq.(128) cannot be achieved without introducing brane bending.  Thus to achieve $h_{55} = 0$, we see from Eqs.(28) and (50) one requires
\begin{equation}
\xi_{5}(x^i,y) = \frac{1}{6A'}f^{T}(x^i,y) + \phi_{5}(x)
\end{equation}
where the function of integration $\phi_{5}(x)$ is independent of y and $f^T(x^i,y)$ is the value of $f^T$ in the frame of Eq.(20).  On the branes, therefore $f^T(y_{\alpha})$ is given by Eq.(58), and one cannot choose $\phi_{5}(x)$ to make $\xi_5$ vanish on both branes.  Thus if we choose $\phi_{5}(x)$ = -$f^T(x^i,y_1)/6A'$ (so that $\xi_5(y_1) = 0$) then $\xi_5(y_2)$ is proportional to $e^{2\beta y_2}\bar{m}(y_2)$ [by Eqs.(58) and (117)] and so there is a huge amount of brane bending on the TeV brane.  (Alternately, the choice $\phi_5(x)$ = 0 gives by Eq.(30) that $f^T(x^i,y)$ = 0 but with brane bending on both branes.)  This would presumably greatly modify the geodesic motion of particles on the TeV brane.  Ref.\cite{callin} carries out the analysis in the frame of Eq.(128) assuming there is no brane bending in that frame.  They define the gravitational potential by the diagram of two point mass stress tensors connected by a free field gravitational propagator.  (The $f^T$ components will vanish for free fields.)  However, to get a consistent solution they find it necessary to fine tune the matter on the $y_2$ brane.  In contrast, the analysis given here is valid for arbitrary matter on the $y_1$ and $y_2$ branes.  Finally we note that none of the previous discussions have analysed gravitational forces involving particles on the $y_2$ brane which is where difficulties arose.

\section*{Note Added}

After the completion of this work, it was brought to our attention that there exists a previous work that examines the corrections to Newton's law in a coordinate frame where $h_{55} \neq 0$ and the branes are kept unbent \cite{smolyakov}.  However we disagree with their results for the Newtonian potential.

\section{Acknowledgments}
This work is supported in part by a National Science Foundation Grant
PHY-0101015

\appendix
\section{Appendix}

\renewcommand{\theequation}{\Alph{section}.\arabic{equation}}
\setcounter{equation}{0}

In Sec.4 we calculated the Kaluza-Klein (KK)corrections to $h_{00}^{TT}$ on the $y_1 = 0$ brane in the case where both particles reside on the $y_1 = 0$ brane.  In this Appendix we will calculate the other KK corrections for the two cases (i) $\xi_1 \ll 1$, $\xi_2 \gg 1$ and (ii) $\xi_1 \gg 1$, $\xi_2 \gg 1$.  These can be most easily found in terms of what we have already shown for $h_{00}^{TT}(y_1)$ due to the presence of $T_{00}(y_1)$.  These results were
\begin{equation}
h_{00}^{TT\,\,(i)}(1,1) = \sum_{n}\frac{R_{n}^{i}(1,1)}{m^2 - m_{n}^2}T_{00}(1)\,\,\,;\,\,\,R_{n}^{i}(1,1) = -\frac{2\pi e^{-\beta y_2}}{3M_{5}^3}m_n
\end{equation}
and
\begin{equation}
h_{00}^{TT\,\,(ii)}(1,1) = \sum_{n}\frac{R_{n}^{ii}(1,1)}{m^2 - m_{n}^2}T_{00}(1)\,\,\,;\,\,\,R_{n}^{ii}(1,1) = -\frac{4\beta e^{-\beta y_2}}{3M_{5}^3}m_n
\end{equation}
where we have denoted the contribution to $h_{00}^{TT}$ on the i'th brane due to matter on the j'th brane by $h_{00}^{TT}(i,j)$ and (i), (ii) represent the two limits on $\xi_1$ and $\xi_2$ stated above. 

We can convert these into coordinate space using
\begin{equation}
m^2 = -p^2 \cong -\vec{p}^{\,2} 
\end{equation}
where the last approximation is true in the static limit.  We then take the continuum limit where
\begin{equation}
\Delta m_n \rightarrow dm_n = \pi\beta e^{-\beta y_2}
\end{equation}
  Thus for Eq.(A.1)
\begin{equation}
h_{00}^{TT\,\,(i)}(1,1) = \frac{2}{3M_{5}^{3}\beta}\int_{0}^{\beta}dm_{n}d^{3}r\frac{e^{ip\cdot r}m_n}{\vec{p}^{\,2} + m_{n}^2}T_{00}(1)
\end{equation}
where we have taken the upper limit of the integral to be $\beta$ since $\xi_1 \lesssim$ 1.  After performing the coordinate space integral we get
\begin{equation}
h_{00}^{TT\,\,(i)}(1,1) = \frac{2}{3M_{5}^{3}\beta}\int_{0}^{\beta}dm_{n}\frac{m_n e^{-m_n r}}{4\pi r}T_{00}(1)
\end{equation}
Upon integration we find
\begin{equation}
h_{00}^{TT\,\,(i)}(1,1) = \frac{1}{6\pi M_{5}^{3}\beta r^3}[1- e^{-\beta r}(\beta r + 1)]T_{00}(1)
\end{equation}
Thus in the limit $r \gg 1/\beta$ we have a $1/r^3$ correction
\begin{equation}
h_{00}^{TT\,\,(i)}(1,1) \cong \frac{1}{6\pi M_{5}^{3}\beta r^3}T_{00}(1)\,\,\,;\,\,\, \xi_1 \ll 1\,;\,\xi_2 \gg 1
\end{equation}
Similarly from Eq.(A.2) we get
\begin{equation}
h_{00}^{TT\,\,(ii)}(1,1) = \frac{1}{3\pi^{2}M_{5}^{3}r^2}e^{-\beta r}T_{00}(1)
\end{equation}
which in the limit $r \ll 1/\beta$ becomes a $1/r^2$ correction
\begin{equation}
h_{00}^{TT\,\,(ii)}(1,1) \cong \frac{1}{3\pi^{2}M_{5}^{3}r^2}T_{00}(1)
\end{equation}

We consider now the other corrections arising from Eqs.(88) and(89).  We have for the correction due to $T_{00}(2)$ on the $y_1$ brane
\begin{equation}
h_{00}^{TT}(1,2) = \frac{\xi_{1}J_{1}(\xi_{1})T_{00}(2)}{\xi_{2}J_{1}(\xi_{2})T_{00}(1)}h_{00}^{TT}(1,1)
\end{equation}
For case (i) we have
\begin{equation}
\frac{\xi_{1}J_{1}(\xi_{1})}{\xi_{2}J_{1}(\xi_{2})} \cong e^{-\beta y_2}\frac{\xi_1}{2\sqrt{\frac{2}{\pi\xi_2}}cos(\xi_2 - \frac{3\pi}{4})}
\end{equation}
From Eq.(90) $tan(\xi_2 -\frac{3\pi}{4}) = -4/(\pi\xi_{1}^2)$ and so
\begin{equation}
cos(\xi_2 - \frac{3\pi}{4}) = \frac{1}{\sqrt{1 + \frac{16}{\pi^{2}\xi_{1}^4}}} \cong \frac{\pi\xi_{1}^2}{4}
\end{equation}
Thus
\begin{equation}
\frac{\xi_{1}J_{1}(\xi_{1})}{\xi_{2}J_{1}(\xi_{2})} =(\frac{2\beta}{\pi m_n})^{1/2}e^{-\frac{\beta y_2}{2}}
\end{equation}
Substituting this expression into Eq.(A.11) and after performing the coordinate space integration we find
\begin{equation}
h_{00}^{TT\,\,(i)}(1,2) = \frac{e^{-\frac{\beta y_2}{2}}}{6\pi^{3/2}M_{5}^{3}r}\sqrt{\frac{2}{\beta}}\int_{0}^{\beta}dm_{n}m_{n}^{1/2}e^{-m_n r}T_{00}(2)
\end{equation}
We can calculate the integral in the limit $r \gg 1/\beta$ which gives a $1/r^{5/2}$ correction
\begin{equation}
h_{00}^{TT\,\,(i)}(1,2) = \frac{\sqrt{2}e^{-\frac{\beta y_2}{2}}}{12\pi M_{5}^{3}}\frac{1}{\beta^{1/2}r^{5/2}}T_{00}(2)
\end{equation}
For case (ii) we need
\begin{equation}
\frac{\xi_{1}J_1(\xi_1)}{\xi_{2}J_1(\xi_2)} \cong \sqrt{\frac{\xi_1}{\xi_2}}\frac{cos(\xi_1 - \frac{3\pi}{4})}{cos(\xi_2 - \frac{3\pi}{4})} = e^{-\frac{\beta y_2}{2}}(-1)^n
\end{equation}
After substituting this into Eq.(A.11) and performing the coordinate space integral we find
\begin{equation}
h_{00}^{TT\,\,(ii)}(1,2) = \frac{\beta}{3\pi M_{5}^3}\frac{e^{-\frac{3\beta y_2}{2}}}{r}\sum_{n}(-1)^{n}e^{-m_n r}T_{00}(2)
\end{equation}
Here since $\xi_1 \gtrsim 1$ the sum is over $m_n \gtrsim \beta$ and since $m_n = n\pi\beta e^{-\beta y_2}$ we require
\begin{equation}
n \gtrsim \frac{e^{\beta y_2}}{\pi} \equiv N \gg 1
\end{equation}
Thus Eq.(A.18) becomes
\begin{equation}
h_{00}^{TT\,\,(ii)}(1,2) \cong \frac{\beta}{3\pi M_{5}^3}\frac{e^{-\frac{3\beta y_2}{2}}}{r}\sum_{n=N}^{\infty}(-1)^{n}e^{-n\pi\beta r e^{-\beta y_2}}T_{00}(2)
\end{equation}
Let 
\begin{equation}
n = N + m \,\,\,;\,\,\, m = 0,1,2,...
\end{equation}
Since 
\begin{equation}
N\pi\beta r e^{-beta y_2} = \beta r
\end{equation}
one has
\begin{equation}
h_{00}^{TT\,\,(ii)}(1,2) \cong (-1)^{N}\frac{\beta}{3\pi M_{5}^3}\frac{e^{-\frac{3\beta y_2}{2}}e^{-\beta r}}{r}\sum_{m=0}^{\infty}(-1)^{m}e^{-m\pi\beta r e^{-\beta y_2}}T_{00}(2)
\end{equation}
The sum is found to give
\begin{equation}
\sum_{m=0}^{\infty}(-1)^{m}e^{-m\pi\beta r e^{-\beta y_2}}T_{00}(2) = \frac{T_{00}(2)}{1 -e^{-\pi\beta r e^{-\beta y_2}} }
\end{equation}
Thus in the limit $r \ll 1/\beta$
\begin{equation}
h_{00}^{TT\,\,(ii)}(1,2) = (-1)^{N}\frac{e^{-\frac{\beta y_2}{2}}}{3\pi^2 M_{5}^3 r^2}T_{00}(2)
\end{equation}

For the KK corrections on the $y_2$ brane we have
\begin{equation}
h_{00}(2,1) = e^{2\beta y_2}h_{00}(1,2)\frac{T_{00}(y_1)}{T_{00}(y_2)}
\end{equation}
Thus from Eqs.(A.16) and (A.25) we find
\begin{eqnarray}
h_{00}^{TT\,\,(i)}(2,1) = \frac{\sqrt{2}e^{\frac{3\beta y_2}{2}}}{12\pi M_{5}^{3}}\frac{1}{\beta^{1/2}r^{5/2}}T_{00}(1)\,\,\,; \beta r \gg 1\\
h_{00}^{TT\,\,(ii)}(2,1) = (-1)^{N}\frac{e^{\frac{3\beta y_2}{2}}}{3\pi^2 M_{5}^3 r^2}T_{00}(1)\,\,\,; \beta r \ll 1
\end{eqnarray}
Similarly the corrections for both particles on the $y_2$ brane give
\begin{equation}
h_{00}^{TT}(2,2) = h_{00}^{TT}(1,1)\frac{J_{1}^2(\xi_1)T_{00}(2)}{J_{1}^2(\xi_2)T_{00}(1)}
\end{equation}
For case (i) we need
\begin{equation}
\frac{J_{1}^2(\xi_1)}{J_{1}^2(\xi_2)} \cong \frac{m_n e^{\beta y_2}}{2\beta}
\end{equation}
Hence
\begin{equation}
h_{00}^{TT\,\,(i)}(2,2) = \frac{\pi}{3\beta M_{5}^3}\sum_{n}\frac{m_{n}^2}{\vec{p}^{\,2} + m_{n}^2}T_{00}(2)
\end{equation}
and after going to the continuum limit and performing the integrations over coordinate space and the mass spectrum we find in the limit $\beta r \gg 1$
\begin{equation}
h_{00}^{TT\,\,(i)}(2,2) = \frac{e^{\beta y_2}}{6\pi\beta^{2}M_{5}^{3}r^4}T_{00}(2)
\end{equation}
For case (ii) we have
\begin{equation}
\frac{J_{1}^2(\xi_1)}{J_{1}^2(\xi_2)} \cong \frac{\xi_{2}cos^{2}(\xi_1 - \frac{3\pi}{4})}{\xi_{1}cos^{2}(\xi_2 - \frac{3\pi}{4})} = e^{\beta y_2}
\end{equation}
which in the limit $\beta r \ll 1$ gives
\begin{equation}
h_{00}^{TT\,\,(ii)}(2,2) = \frac{e^{\beta y_2}}{3\pi^{2}M_{5}^{3}r^2}T_{00}(2)
\end{equation}


\begin{thebibliography}{99}
\bibitem{ibanez}L.E. Ib\'a\~nez, [hep-ph/0109082].
\bibitem{hw}P. Horava and E. Witten \Journal{\NPB}{460}{506-524}{1996}[hep-th/9510209].
\bibitem{hw2}P. Horava and E. Witten, \Journal{\NPB}{475}{94-114}{1996} [hep-th/9603142].
\bibitem{witten}E. Witten, \Journal{\NPB}{471}{135-158}{1996} [hep-th/9602070].
\bibitem{horava}P. Horava, \Journal{\PRD}{54}{7561-7569}{1996} [hep-th/9608019].
\bibitem{rs}L. Randall and R. Sundrum,\Journal{\PRL}{83}{3370-3373}{1999} [hep-ph/9905221].
\bibitem{rs2}L. Randall and R. Sundrum,\Journal{\PRL}{83}{4690-4693}{1999} [hep-th/9906064].
\bibitem{binetruy}P. Binetruy, C. Deffayet, and D. Langlois, \Journal{\NPB}{565}{269-287}{2000} [hep-th/9905012].
\bibitem{csaki}C. Csaki, M. Graessner, L. Randall, and J. Terning, \Journal{\PRD}{62}{045015}{2000} [hep-ph/9911406].
\bibitem{khoury}J. Khoury and R. Zhang, \Journal{\PRL}{89}{061302}{2002} [hep-th/0203274].
\bibitem{dewolfe}O. DeWolfe, D.Z. Freedman, S.S. Gubser, and A. Karch, \Journal{\PRD}{62}{046008}{2000} [hep-th/9909134].
\bibitem{cline}J. Cline and H. Firouzjahi, \Journal{\PLB}{495}{271-276}{2000} [hep-th/0008185].
\bibitem{cline2}J. Cline and H. Firouzjahi, \Journal{\PRD}{64}{023505}{2001} [hep-ph/0005235].
\bibitem{enq}K. Enqvist, E Keski-Vakkuri, and S. R$\ddot{a}$s$\ddot{a}$nen, \Journal{\NPB}{614}{388-401}{2001} [hep-th/0106282].
\bibitem{ellwanger}U. Ellwanger, \Journal{\EJC}{25}{157-164}{2002}[hep-th/0001126].
\bibitem{lukas}A. Lukas and D. Skinner, \Journal{\JHEP}{0109}{020}{2001}[hep-th/0106190].
\bibitem{kobayashi}Kobayashi, \Journal{\JHEP}{0212}{056}{2002} [hep-th/0210029].
\bibitem{add}R. Arnowitt, J. Dent, and B. Dutta [hep-th/0405050].
\bibitem{gw}W. Goldberger and M. Wise, \Journal{\PRL}{83}{4922-4925}{1999} [hep-ph/9907447].
\bibitem{lykken}J. Lykken and L. Randall, \Journal{\JHEP}{0006}{014}{2000} [hep-th/9908076].
\bibitem{tanaka}J. Garriga and T. Tanaka \Journal{\PRL}{84}{2778-2781}{2000}[hep-th/9911055].
\bibitem{giddings}S. Giddings, E. Katz, and L. Randall \Journal{\JHEP}{0003}{23}{2000}[hep-th/0002091].
\bibitem{chung}D.J.H. Chung, L. Everett, and H. Davoudiasl \Journal{\PRD}{64}{065002}{2001}[hep-ph/0010103].
\bibitem{dorca}M. Dorca and C. van de Bruck \Journal{\NPB}{605}{215-233}{2001}[hep-th/0012116].
\bibitem{deruelle}N. Deruelle and T. Dolezel \Journal{\PRD}{64}{103506}{2001}[gr-qc/0105118].
\bibitem{nojiri}S. Nojiri and S. Odintsov \Journal{\PLB}{548}{215-223}{2002}[hep-th/0209066].
\bibitem{callin}P. Callin and F. Ravndal [hep-ph/0403302].
\bibitem{adm}R. Arnowitt, S. Deser, and C. Misner \Journal{Phys.Rev}{117}{1595}{1960}.
\bibitem{adm2}R. Arnowitt, S. Deser, and C. Misner [gr-qc/0405109].
\bibitem{smolyakov}M.N. Smolyakov and I.P. Volobuev \Journal{\TMP}{139}{458}{2004}[hep-th/0208025].


\end{thebibliography}
\end{document}